\documentclass[twocolumn,showpacs,showkeys,preprintnumbers,amsmath,amssymb,superscriptaddress]{revtex4}

\usepackage{graphicx}% Include figure files
\usepackage{dcolumn}% Align table columns on decimal point
\usepackage{bm}% bold math

\def\bea{\begin{eqnarray}}
\def\eea{\end{eqnarray}}

\begin{document}

\preprint{Version 1.2}

\keywords{fluctuations, autocorrelations, inverse problem, heavy ions, relativistic nuclear collisions}

\title{Autocorrelations from fluctuation scale dependence by inversion}

\author{T.\,A.~Trainor, R.\,J.~Porter and D.\,J.~Prindle \\
{\em \small CENPA 354290 University of Washington, Seattle, WA 98195}}

%\address{CENPA 354290 University of Washington, Seattle, WA 98195}

\date{\today}

\begin{abstract}
Fluctuations in nuclear collisions can be measured as a function of momentum-space binning scale over a scale interval bounded by detector two-track resolution and acceptance. Fluctuation scale dependence is related to two-particle correlations by a Fredholm integral equation. That equation can be inverted by standard numerical methods to yield an autocorrelation distribution on difference variables as a projection of the full two-particle distribution which retains most of the correlation information in a more compact form. Autocorrelation distributions are typically more easily interpreted in terms of physical mechanisms than fluctuation measurements. \\ \\
\end{abstract}
%\twocolumn

\pacs{24.60.Ky,25.75.Gz}

\maketitle

\section{Introduction}

The dynamics of relativistic nuclear collisions are studied {\em via} the momentum structure of the hadronic final state~\cite{corrsurv}. The correlation structure of pair density $\rho(\vec{p}_1,\vec{p}_2)$ on two-particle momentum space provides qualitatively new information about collision dynamics compared to single-particle density $\rho(\vec{p})$. Since the full two-particle momentum space cannot be visualized, it is necessary to project the pair distribution onto subspaces, retaining as much of the correlation structure as possible in the projections. For relativistic nuclear collisions the {\em autocorrelation} distribution~\cite{autocorr} represents a nearly lossless projection~\cite{axialcd} of the two-particle momentum distribution onto visualizable subspaces ({\em i.e.}, projections onto two dimensions). 
  
Measurements of the fluctuations of event-wise multiplicity $n$ and mean transverse momentum $\langle p_t \rangle$ have been proposed to study the QGP phase boundary as probed by heavy ion collisions~\cite{Rstock,Gbaym,Shuryak,SRS,Berdnikov}. Multiplicity and mean-$p_t$ fluctuations may also result from hard parton scattering and the hadronization process~\cite{hijsca,axialcd}. Direct physical interpretation of fluctuation measurements is typically difficult. However, the bin-size or {\em scale} dependence of fluctuations is determined by the correlation structure of the underlying distribution~\cite{cltps,ptprl}. We have determined that fluctuation scale dependence can be inverted by a numerical procedure to obtain autocorrelation projections which can be interpreted directly. %The inversion technique provides direct correspondence of fluctuation measurements with physical mechanisms.

In this paper we derive an exact relation between fluctuation scale dependence and autocorrelation distributions in the form of an integral equation. We describe the procedure required to invert the integral equation~\cite{invtheor} and thereby obtain autocorrelations from fluctuation scale dependence. We apply that procedure to Monte Carlo simulations and examine the relation of the revealed correlations to known Monte Carlo physics and the precision of the autocorrelations obtained by inversion compared to direct pair analysis.%, demonstrating that the resulting autocorrelations are interpretable in terms of physical mechanisms.

\section{Previous Work}

\subsection{Scaled factorial moments}

Intermittency analysis
% was motivated by observation of apparently anomalous statistical fluctuations in a small number of collision events~\cite{jaycee}. That analysis proposal 
introduced the concept that fluctuation dependence on bin size or scale should be examined over a range of scales to study particle production mechanisms in nuclear collisions~\cite{bialas}. The principal motivation was the conjecture that some particle production mechanisms may produce fluctuations which saturate at a particular scale, whereas other mechanisms ({\em e.g.,} random cascades) may exhibit power-law behavior limited only by finite experimental resolution. The proposal confronted the problem of detecting fluctuation {\em excess} (beyond that due to finite particle number) by introducing {\em normalized factorial moments} which take on unit value for Poisson fluctuations. Power-law (fractal) behavior is expected theoretically from both QCD branching processes and some models of the QCD phase transition ({\em e.g.,} critical fluctuations), typically motivating log-log plots of normalized factorial moments {\em vs} binning scale. A major experimental program to detect and study power-law behavior based on factorial moment analysis has met with considerable success~\cite{corrsurv}.

%Average over events and bins (cells), horizontal and vertical. The difference is the form of the reference. For the horizontal, the bin average over the acceptance is the reference, so that mean distribution structure is included in the factorial moment information.

\subsection{Correlation integrals}

The correlation integral~\cite{corrint}, introduced as a supplement to factorial moment analysis, establishes the relation between moments (ordinary moments, factorial moments, cumulants) of a multiparticle density distribution and integrals of that distribution over certain momentum subspaces. The integrand of a correlation integral can be expressed as a projection of the full multiparticle space onto a subspace of momentum difference variables. If that projection is obtained by averaging it is an autocorrelation. A connection is thus established between running integrals of autocorrelations on difference variables and scale dependence of distribution moments.

%Rank-2 cumulant correlation function:
%$C_2 \equiv \rho(p_1,p_2) - \rho(p_1)\rho(p_2)$ factorization - vs mixed pairs
%Alternatively, normalized correlations $\rho(p_1,p_2) / \rho(p_1)\rho(p_2)$. Claim that $q \ll \bar n$ is required. Not true.

\subsection{Heavy ions and the integral equation}

%The unnormalized factorial moments are defined by $\tilde F_q = \langle n^{[q]} \rangle \equiv \langle n!/(n-q)! \rangle$. Normalized is $F_q \equiv \hat F_q / \langle n \rangle^q$, a particular choice of reference: Poisson fluctuations. Presumption of {\em scaling laws} $F_q(\delta y) \propto \delta y^{-\alpha_q}$. That expectation naturally suggest a logarithmic plotting format.

Normalized factorial moments plotted on a log format %(representing {\em differences} between object and reference distributions) 
{\em vs} logarithmic scale variables have been broadly applied to elementary collisions. In a heavy ion context the analysis problem changes significantly. The accessible scale interval is typically small -- 1-2 decades, but the statistical power of the data is typically very large and the scale dependence may be highly structured over small scale intervals (as opposed to a fixed power-law  trend over a large interval). The ensemble-averaged distribution providing the reference may be complex, and strongly dependent on collision parameters such as energy and centrality. Thus, the advantage of normalized factorial moments in terms of the Poisson reference does not offset difficulties with low-multiplicity data (empty-bin effect). 

Those features lead in heavy ion analysis to three modifications of the factorial-moment analysis: 1) ordinary moments replace factorial moments, 2) the differential fluctuation measure is a linear combination of moments ({\em e.g.,} variance or higher cumulant differences) with the cumulant reference provided by local bin averages (as for a `vertical' factorial moment analysis) and 3) the scale dependence is plotted on a linear format -- {\em e.g.,} linear variance difference {\em vs} linear bin size. That linear framework combined with the correlation integral concept leads to the definition of an invertible integral equation connecting fluctuation scale dependence to autocorrelation distributions, as described below. %In small words, if bin sizes change the variances change because of relation of bin size to correlation structure. Small bins do not see large-scale correlations.

\section{Statement of the Problem}

\subsection{Motivation}

Motivations for this work are three-fold: 1) Fluctuation measurements have been promoted to study branching processes (intermittency, QCD bremsstrahlung) and critical charge, flavor and isospin fluctuations associated with the QCD phase boundary. Interpretation of fluctuation measurements is indirect and model dependent, whereas inversion to correlations usually leads to straightforward interpretations; 2) Direct calculation of two-particle correlations is computer intensive -- $O(n^2)$, whereas calculation of fluctuations is fast -- $O(n)$. If correlations are derived from fluctuations considerable time savings are possible, most important for heavy ion data samples of millions of events with thousands of particles; 3) Two-particle correlations computed directly as pair ratios or differences contain an offset ambiguity related to relative normalization of object and reference distributions. The ambiguity is eliminated for correlations obtained by fluctuation inversion.

\subsection{Basic concepts}

The single-particle number density $\rho(\vec{p}) \leftrightarrow  d^3n/dp^3$ with integral $\langle n \rangle$ is {\em estimated} by accumulated particle counts on a binned momentum space (histogram). Histograms for individual events sample the (unobserved) event-wise {\em parent} density distributions determined by collision dynamics. If the parent is the same for all events the event-wise estimators deviate from the ensemble mean only according to Poisson statistics. The ensemble-averaged histogram then contains all information about the parent distribution and the dynamics.

If the parent distribution for individual events deviates significantly from the mean, the variation produces {\em non}-Poisson fluctuations of histogram bin contents. The scale (bin size) for each momentum component determines the amount of information in the fluctuations. In this work fluctuations are measured by variances computed for bins over a range of scales. The scale dependence of variances reveals the spatial structure of event-wise changes of the parent distribution (as outlined in~\cite{bialas} for factorial moments), and therefore the collision dynamics.

%Difference histogram estimates exact correlation probability rho12 - rho1*rho2, which is the connected two-point correlation function.

The two-particle pair-number density $\rho(\vec{p}_1,\vec{p}_2) \leftrightarrow  d^6n/dp^3_1dp^3_2$ with integral $\langle n(n-1) \rangle$ is similarly estimated by pair counts in a histogram on the two-particle momentum space. Just as a single-particle histogram has a mean-value component and possibly non-Poisson bin fluctuations about the mean, the two-particle distribution has similar components. However, the {\em mean values} of the two-particle histogram are related to the {\em fluctuations} of the single-particle distribution, suggesting an algebraic relationship between two-particle correlations and single-particle fluctuations.

\subsection{Correlations and autocorrelations}

Two-particle correlations are measured by the difference between object and reference histograms. The object distribution is the ensemble mean of event-wise two-particle histograms (sibling pairs). The reference distribution is either the cartesian product of single-particle distributions or a histogram of particle pairs formed from different events (mixed pairs). In either case the reference  represents particles produced as independent samples from the single-particle parent. The object contains additional information on correlated particle production, for instance emission from several relatively-moving sources.

Two-particle difference histograms on momentum components pseudorapidity $\eta$ and azimuth $\phi$ are observed to be approximately invariant along the sum axis ({\em e.g.,} $\eta_1 + \eta_2$) for relativistic nuclear collisions. Such invariance is consistent with two possibilities: 1) the correlation mechanism is truly invariant on one or both variables ({\em e.g.,} hydrodynamic flow) or 2) structure is randomly distributed in the single-particle momentum space from event to event, but produces a structure invariant on the sum variable in the ensemble-averaged histogram ({\em e.g.,} jet correlations). In either case, the structure on the difference variables ({\em e.g.,} $\eta_1 - \eta_2$) then represents all correlation information as a projection by averaging: an {\em autocorrelation}.

An autocorrelation compares a distribution $f(x)$ to itself in the form of an average of product $f(x_1)f(x_2)$ along the sum variable $x_1 + x_2$ for each value of the difference variable $x_1 - x_2$. That projection by averaging onto the difference variable is a {\em lossless} projection strategy if correlations are invariant on the sum variable (stationary). Autocorrelations are well-established in time-series analysis (where the difference variable is the {\em lag} $\tau$)~\cite{auto1}. They also appear in various aspects of nuclear collision analysis. Direct determination of {\em joint} autocorrelations on momentum subspace $(\eta,\phi)$ has been extensively implemented in multiparticle correlation analysis of fixed-target p-p experiments~\cite{multicorr}. Autocorrelations also provide unbiased access to jet correlations, as we demonstrate in the section on Monte Carlo simulations. The examples above involve {\em direct} determination of autocorrelations by pair counting in histograms. Alternatively, autocorrelations can be obtained by inverting the bin-size or scale dependence of fluctuations. 

\subsection{A simple example}

An event-wise parent distribution on $x$ is defined as a gaussian superposed on a flat background. In each `event' the gaussian appears at a random position. The parent for each event is sampled randomly. The ensemble mean sample distribution on $x$ (single-point distribution) is uniform, retaining no information about the gaussian shape. The sample-{\em pair} distribution for each event is histogrammed on $(x_1,x_2)$ (two-point distribution), and the object distribution is the ensemble-average pair distribution, which includes a uniform component from the flat background, a second uniform component from gaussian pairs distributed randomly on $(x_1,x_2)$, and a ridge along the sum diagonal representing sample pairs from the same gaussian (gaussian self pairs -- {\em sample} self pairs are not included in the histogram). The reference distribution formed with sample pairs taken from different events is uniform (cartesian product of single-point distributions). The object-reference difference reveals the ridge along the sum diagonal, which represents the gaussian component of the parent. The projection of that difference histogram onto difference variable $x_1 - x_2$ by {\em averaging along the sum variable} is an autocorrelation. The principal feature of the autocorrelation is a gaussian, with amplitude and width determined by the mean of the parent gaussians.

From the same event ensemble the sample variances from a fluctuation analysis can be obtained from histograms on $x$ of varying bin sizes. The variance difference between object and reference distributions plotted as a function of increasing bin size exhibits a rapid initial rise when the bin size is less than the gaussian width, followed by asymptotic approach to a slower linear increase that continues indefinitely. That fluctuation scale dependence is the integral on the difference variable of the gaussian in the autocorrelation plus the uniform background, and may be processed by the methods described in this paper to infer the average properties of the parent distribution, including the gaussians. 

%This example illustrates how an autocorrelation accesses localized structure randomly occuring in each event but having persistent characteristics over an event ensemble, and how the scale dependence of fluctuations is related to that autocorrelation.

\subsection{Paper outline}

We first derive the exact linear relationship between the bin-size dependence of fluctuations measured in a single-particle space and the integral of an autocorrelation projection of the two-particle difference histogram. We then introduce an inversion method to solve the integral equation so that given the fluctuation information the autocorrelation may be inferred. Finally, we demonstrate physical interpretation of autocorrelations and the precision of the inversion method using Monte Carlo simulations relevant to nuclear collisions.

%%%%%%%%%%%%%%%%%%%%%%

%\section{Autocorrelations}

%A series of correlation measurements on rapidity in elementary collisions $e^+$-$e^-$, $p$-$p$ and $p$-$\bar p$~\cite{epem} was critical to formulating the phenomenological representation of nonperturbative QCD contained in the Lund model~\cite{lund}. To accommodate limited statistics the two-particle distribution on rapidity was typically divided into two regions by conditions placed on the rapidity of one (test or tagged) particle. The joint distribution for particles of opposite `charge' was then projected onto the rapidity of a second particle. Those conditional projections approximate an autocorrelation distribution. 

%HBT analysis of quantum correlations in nuclear collisions also employs autocorrelations on momentum difference variables~\cite{hbt}.
%~\cite{axialci}. 

%Inversion of fluctuation scale dependence to obtain autocorrelations provides three advantages: 1) direct determination of an autocorrelation by pair ratios is $O(n^2)$, whereas a fluctuation calculation is $O(n)$, thereby reducing computation time; 2) autocorrelations obtained directly from pair ratios may have an arbitrary offset due to pair normalization ambiguities; and 3) as noted, fluctuations are less easily interpreted than autocorrelations in terms of physical mechanisms.

\section{Summary of the Method}

We introduce the basic concepts of the paper and define the integral equation between fluctuation scale dependence and two-particle correlations in the form of autocorrelation distributions.

\subparagraph{Spaces and measures}

The particle momentum in a relativistic nuclear collision is expressed as $\vec{p}\rightarrow (p_t,\eta,\phi)$, where $p_t$ is the transverse momentum relative to the collision axis $\hat z$, $\eta \equiv \frac{1}{2}\ln\{\frac{p+p_z}{p - p_z}\}$
 is the pseudorapidity and $\phi$ is the momentum azimuth angle.
%, $y_t \equiv \ln\{(m_t + p_t)/m_0\}$ is transverse rapidity (longitudinally co-moving frame $p_z = 0$, $m_t^2 \equiv p_t^2 + m_0^2$) and $y_z \equiv \ln\{(E + p_z)/m_t\}$ is longitudinal rapidity. Mass $m_0$ is assigned as the pion mass in the absence of particle identification. 
Given two-particle momentum space $(\vec{p}_1,\vec{p}_2)$, each momentum component $x$, bounded by some detector acceptance $\Delta x$, defines a two-particle subspace $(x_1,x_2)$. To accommodate numerical integration we define a {\em two-tiered} binning system including microbins and macrobins. Subspace $x$ is partitioned into {\em micro}bins of {\em fixed size} $\epsilon_x$ (limited below by detector resolution). {\em Macro}bins, with variable bin size or scale $\delta x$ such that $\epsilon_x \leq \delta x \leq \Delta x$, are defined as combinations of the underlying microbins. Variation of fluctuations with macrobin scale is the basis of the fluctuation inversion analysis. 

If a measure $W$ is distributed on $x$ the corresponding histogram bin contents are denoted $w$ ({\em e.g.,} scalar sum of particle $p_t$ or particle multiplicity $n$ in a bin). We refer to single-particle measure density $\rho(w;x)$ and histogram bin contents $\bar w_a \simeq \epsilon_x \rho(w;x_a)$, or two-particle density $\rho(w;x_1,x_2)$ and histogram $\overline{w_a\,w_b} \simeq \epsilon^2_x\rho(w;x_a,x_b)$. An object histogram is formed with sibling pairs taken from same events. The reference histogram is formed with mixed pairs taken from different events. Given object distribution $\rho_{obj}$ and reference distribution $\rho_{ref}$ the difference density (correlation function) $\Delta \rho \equiv \rho_{obj} - \rho_{ref}$ represents the desired correlation structure, and $\Delta w \equiv w_{obj} - w_{ref}$ represents the corresponding difference histogram. 

\subparagraph{Fluctuations}

%Excess fluctuations are determined by correlations.

%Fluctuations are determined by the correlation structure of binned distributions. 
In a nuclear collision a total weight $W$ ({\em e.g.,} total transverse momentum $P_t$ or multiplicity $N$) falls within a detector acceptance $(\Delta \eta,\Delta \phi)$ divided into $M$ {macro}bins of size $(\delta \eta,\delta \phi)$. Each macrobin (represented by single scale $\delta x$ for brevity) contains in each event some integrated weight $w(\delta x)$ as a random variable. The conventional per-bin variance is $\sigma_{w}^2(\delta x) \equiv \overline{\{w(\delta x)-\bar w(\delta x)\}^2}$, where the overlines represent  averages over all macrobins of size $\delta x$ in all events of an ensemble~\cite{ptprl}. The corresponding scale-dependent {\em per-particle} variance is defined by $\sigma_{w/}^2(\delta x) \equiv \sigma_{w}^2(\delta x)/\bar n(\delta x)$. Its small-scale limit $\sigma^2_{\hat w}$ ($\bar n = 1$) is ensemble-mean single-particle $p_t$ variance $\sigma^2_{\hat p_t}$ if $w = p_t$, and is defined to be 1 (Poisson limit) if $w = n$. The scale-dependent {\em variance difference} defined by $\Delta \sigma_{w/}^2(\delta x) \equiv \sigma_{w/}^2(\delta x) - \sigma^2_{\hat w}$ is a definite integral of two-particle correlations over a scale interval~\cite{cltps}, as demonstrated below.

\subparagraph{Autocorrelations}

%Autocorrelation $A$ is the projection by averaging of a two-particle distribution onto difference variables. 
For distribution $f(x)$ the autocorrelation is $A(x_\Delta) \equiv \overline{f(x)\cdot f(x + x_\Delta)}$, where the overline denotes an average over some bounded interval (acceptance) on $x$. For relativistic nuclear collisions the 2D {\em joint} autocorrelation on $(\eta_\Delta,\phi_\Delta)$ is a nearly lossless projection from the 4D $(\eta_1,\eta_2,\phi_1,\phi_2)$ pair-momentum space. 
%Projection from space $(p_{t1},p_{t2})$ onto its difference variable is not useful due to strong variation of the pair density on the single-particle $p_t$ space. However, per-particle correlation measure ${\Delta \rho }/{ \sqrt{\rho_{ref}}}$ can also be defined on  $(y_{t1},y_{t2})$, may be nearly stationary over a significant interval on sum variable $y_{t\Sigma}$, and is therefore suitable for projection onto difference variable $y_{t\Delta}$. 
We distinguish between autocorrelation histograms $A$ and autocorrelation densities $\rho(x_\Delta)$. We define histogram difference $\Delta A \equiv A_{obj} - A_{ref}$ and corresponding density difference $\Delta \rho = \rho_{obj} - \rho_{ref}$.

\subparagraph{Integral equation}

Variation of $\Delta \sigma^2_{w/}$ on bin scales $(\delta \eta,\delta \phi)$ is represented by an {\em integral equation} which can be inverted to obtain an autocorrelation on difference variables $(\eta_\Delta,\phi_\Delta)$. The integral equation is 
\bea \label{inverse}
\Delta \sigma^2_{w/}(m \, \epsilon_\eta, n \, \epsilon_\phi) &\equiv& 2 \sigma_{\hat w} \Delta \sigma_{w/}(m \, \epsilon_\eta, n \, \epsilon_\phi) \\ \nonumber
= 4 \sum_{k,l=1}^{m,n} \epsilon_\eta \epsilon_\phi & &\hspace{-.25in} K_{mn;kl}  \,   \frac{\Delta \rho(w;k\,\epsilon_\eta, l\, \epsilon_\phi) }{ \sqrt{\rho_{ref}(n;k\,\epsilon_\eta, l\, \epsilon_\phi)}} ,
\end{eqnarray}
which also defines {\em difference factor} $\Delta \sigma_{w/}$. Derivation of that integral equation and the techniques required for its inversion are the main subjects of this paper. 

%Autocorrelation density ratio ${\Delta \rho }/{ \sqrt{\rho_{ref}}}$ can be directly and often simply interpreted in terms of physical mechanisms. 

\section{Fluctuation Scale Dependence}

%Variance as a fluctuation measure compares an object distribution to a reference composed of averages of the object distribution. The scale (bin size) dependence of fluctuations as measured by a variance is equivalent to a running integral of two-particle correlations. The difference between fluctuations across a scale interval is equivalent to a definite integral of correlations over that scale interval.

%\subparagraph{Measures and binnings}

%The single-particle density $ \rho(w;x)$ of measure $w$ is distributed on bounded 1D space $x \in [-\Delta x/2,\Delta x/2]$, {\em i.e.,} within `acceptance' $\Delta x$. The space is partitioned into equal {macro}bins of size $\delta x$ referred to as the binning {\em scale}. The contents of bin $s$ is $w_s \approx \delta x\, \rho(w,x_s)$. For a comparison of object and reference distribution we define difference histogram $\Delta w(\delta x) \equiv w_{obj}(\delta x) - w_{ref}(\delta x)$.

\subparagraph{Variances}

We define the scale-dependent {\em total variance} of difference histogram $\Delta w(\delta x)$ by
\bea \label{totvar}
 \Sigma^2_{w}(\Delta x,\delta x) \hspace{-.05in}  &\equiv& \hspace{-.15in} \overline
{\sum_{s=1}^{M(\Delta x,\delta x)}
\{\Delta w_{s}(\delta x) \}^2}, 
\eea
where $s$ is a macrobin index, $M(\Delta x,\delta x)$ is the event-wise number of {\em occupied} macrobins of size $\delta x$ in acceptance $\Delta x$, and the bar denotes an average over all events. This quantity is also a bin-based estimate of the {\em correlation integral} $C_2(w;\Delta x,\delta x)$~\cite{corrint}. Because the total variance is extensive with respect to detector acceptance $\Delta x$ we define several intensive forms which result in, respectively, per-bin and per-particle variances and a total-variance {\em density} on $x$
\bea
\Sigma^2_w(\Delta x,\delta x) / M(\Delta x,\delta x)&\equiv& \sigma^2_{w}(\delta x) \\ \nonumber
\Sigma^2_w(\Delta x,\delta x) /\bar{N}(\Delta x) & \equiv &  \overline{   (\Delta w )^2 }/\bar n\\ \nonumber
& \equiv & \sigma^2_{w/}(\delta x) \\ \nonumber
\Sigma^2_w(\Delta x,\delta x) / \Delta x & \simeq & d\bar n/dx \, \sigma^2_{w/}(\delta x).
\eea
Those factorized forms separate acceptance $\Delta x$ and scale $\delta x$ dependencies. The optimum choice among them depends on the physical source(s) of nontrivial fluctuations. 

\subparagraph{Variance differences}

In the small-scale limit where the particle occupancy of all {\em occupied} bins is 1 we have $\Sigma^2_w(\Delta x,\delta x \ll \Delta x) \rightarrow \bar{N}(\Delta x) \, \sigma^2_{\hat w}$, the inclusive single-particle variance of $w$. For {\em central limit conditions} (independent $w$ samples from a fixed parent distribution) the total variance is scale invariant: $\Sigma^2_w(\Delta x,\delta x) \equiv \bar{N}(\Delta x) \, \sigma^2_{\hat w}$ over any scale interval for which CLT conditions are valid. The scale invariance of total variance is an alternative statement of the central limit theorem~\cite{cltps}. Deviations of $\Sigma^2_w(\Delta x,\delta x)$ from CLT invariance indicate the presence of $w$ correlations. That connection leads to the definition of a fluctuation measure which is differential across a scale interval: the {\em total-variance difference}
\bea
  \Delta\Sigma^2_{w}(\delta x_1,\delta x_2)  &\equiv& 
  \Sigma^2_{w}(\delta x_2) - \Sigma^2_{w}(\delta x_1),
\eea
which is zero if CLT conditions apply over the scale interval. The total variance difference for $\delta x_2 = \delta x$ and $\delta x_1
\ll \Delta x$ ($\bar n = 1$) is given by
\bea  \label{EqA11}
  \Delta\Sigma^2_{w}(\Delta x,\delta x) &=&  \bar{N}(\Delta x)
  \, \left\{ \sigma^2_{w/}(\delta x)
- \sigma^2_{\hat w} \right\}    \nonumber \\
&\equiv& \bar{N}(\Delta x)\, \Delta \sigma^2_{w/}(\delta x), 
\eea
which defines {\em per-particle} variance difference $\Delta \sigma^2_{w/}$.
The variance difference across a scale interval is a measure of correlations in $\rho(w;x)$ integrated over that scale interval; it is equivalent to the integral of a difference autocorrelation across the same scale interval, as described below. From that equivalence we obtain the integral equation relating autocorrelations to fluctuation scale dependence.

\section{Autocorrelations}

The autocorrelation of a two-particle distribution on binned space $(x_1,x_2)$ consists of averages over diagonal regions of the 2D space which then occupy single bins of the autocorrelation histogram on the difference variable. We describe the construction of autocorrelations from a two-particle distribution by two direct methods ({\em i.e.,} by pair counting, not fluctuation inversion).

\subparagraph{Two-particle spaces}

%Single-particle density $ \rho_1(w,x)$ is distributed on bounded 1D space $x \in [-\Delta x/2,\Delta x/2]$, {\em i.e.,} within an `acceptance' $\Delta x$. 

\begin{figure}[h]
\includegraphics[keepaspectratio,width=3.3in]{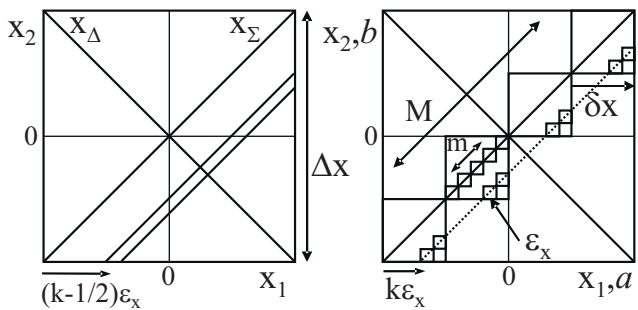}
\caption{Integration schemes for direct construction of autocorrelations: 1) normalized strip integrals (left panel) and 2) diagonal bin averages (right panel). Macrobins have variable size $\delta x$, microbins have fixed size $\epsilon_x$. Bins on difference variable $x_\Delta$ are indexed by integer $k$. \label{bins}}
\end{figure}

In Fig.~\ref{bins}, two-particle density $\rho(w;x_1,x_2)$ is distributed on space $(x_1,x_2)$. It is convenient to define {\em sum and difference variables} $x_\Sigma \equiv x_1 + x_2$ and $x_\Delta \equiv x_1 - x_2$. Distance along those variables is measured on $x$ as shown in Fig.~\ref{bins} to avoid factors $\sqrt{2}$. If the correlation structure of single-particle density $\rho(w;x)$ is independent of mean position on $x$ or {\em stationary} (equivalently, $\rho(w;x_1,x_2) \rightarrow \rho(w;x_\Sigma,x_\Delta)$ is independent of sum variable $x_\Sigma$) the corresponding autocorrelation density on $x_\Delta$ represents a lossless projection (all structure retained) of the two-particle density onto a lower-dimensional space: $\rho(w;x_\Sigma,x_\Delta) \rightarrow \rho(w;x_\Delta)$. For relativistic nuclear collisions that symmetry is nominally exact on $\phi$ (assuming ideal detection) and approximate on $\eta$ (depending on the validity of `boost invariance' over some pseudorapidity interval). 

% Differential correlation measures which compare object and reference distributions are based on density difference distribution $\Delta \rho \equiv \rho_{obj} - \rho_{ref}$. That 2D difference distribution, when binned, is a {\em covariance} distribution.

%%%%%%%%%%%%%%%%%%%%%%%%%%%%%%%%%%
%%%%%%%%%%%%%%%%%%%%%%%%%%%%%%%%%%

\subparagraph{Autocorrelation histograms}

Autocorrelation histograms can be constructed by three methods: 1) more directly (Fig.~\ref{bins}, left panel) by sorting particle pairs into 1D bins on momentum difference variables $x_\Delta$ as normalized strip integrals~\cite{corrint}, 2) less directly (Fig.~\ref{bins}, right panel) as weighted averages of diagonal combinations of microbins from a 2D histogram on space $(x_1,x_2)$, 3) indirectly by {\em inverting} the scale dependence of a moment, variance or variance difference. %Inversion of fluctuation scale dependence to obtain autocorrelations is the subject of this paper. 

\begin{figure}[h]
\includegraphics[keepaspectratio,width=3in]{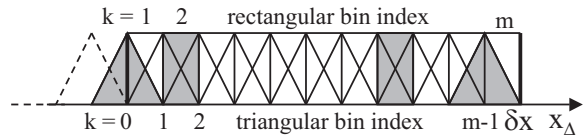}
\caption{Binning schemes on difference variable $x_\Delta$: 1) square bins on $x_\Delta$ (Fig.~\ref{bins} left panel) with $|k| \in [1,m]$, and 2) triangular projections of 2D bins (Fig.~\ref{bins} right panel) with $|k| \in [0,m-1]$. \label{twobins}}
\end{figure}

We introduce for the numerical integration {micro}bins of fixed size $\epsilon_x$ such that there are $m_\delta$ microbins in a {macro}bin of variable size ($\epsilon_x \leq \delta x \leq \Delta x$) and $m_\Delta$ microbins in a fixed acceptance of size $\Delta x$. If difference variable $x_\Delta$ is partitioned into microbins indexed by $k$, two autocorrelation types, which depend on binning strategy on the difference variable, can be defined for an {\em aperiodic} distribution on $x$. Those autocorrelation types correspond to the two direct construction methods described above and illustrated in Fig.~\ref{bins} (left and right panels respectively): 1) square bin edge on zero (strip integral -- particle pairs indexed by $i,j$ falling within the $k^{th}$ $x_\Delta$ bin), then  
\bea \label{auto1}
A^{(1)}_k(w) &\equiv& \frac{1}{m_\Delta - |k| + 1/2} \overline{\sum_{ij=1}^{N} (w_i \, w_{j})_{ij \in k}},
\eea
with $|k| \in [1,m_\Delta]$ ($w_i, w_j$ are single-particle values); and 2) triangular bin centered on zero (averages of 2D microbin contents indexed by $a,b$), then 
\bea  \label{auto2}
A^{(2)}_k(w) \equiv \frac{1}{m_\Delta - |k|} \sum_{a=1}^{m_\Delta-|k|} \overline{w_a \, w_{a+k}},
\eea
with $|k| \in [0,m_\Delta-1]$ (average over all bins on diagonal bin array within acceptance, Fig.~\ref{bins} (right panel) with $\delta x \rightarrow \Delta x$). Overlines indicate averages over all events.  For a distribution {\em periodic}\, on $x$ the type-1 autocorrelation is given by $A^{(1)}_k(w) \equiv 1/m_\Delta \overline{\sum_{ij=1}^{N} (w_i \, w_{j})_{ij \in k}}$, with particles of any pair $i,j$ falling within one period on $x$ (the period then defines $\Delta x$). Binning strategies and indexing on $x_\Delta$ for the two cases are shown in Fig.~\ref{twobins}. For autocorrelations constructed from a particle list (most efficient) the square bin index applies. For prebinned 2D space $(x_1,x_2)$ the triangular bin index applies. For fluctuation inversion leading to autocorrelations either method may be used, but the rectangular bin index facilitates direct comparison between fluctuation inversion and the more efficient direct method 1) of autocorrelation construction.

%%%%%%%%%%%%%%%%%%%%%%%%%%%%%%%%%%
%%%%%%%%%%%%%%%%%%%%%%%%%%%%%%%%%%

\subparagraph{Difference autocorrelations} 

$\Delta A_k = A_{k,obj} - A_{k,ref}$ represents a subset of the correlation content of an {\em object} distribution. The type-1 number $n$ reference autocorrelation obtained with mixed pairs (particles selected from different but similar events) is $A^{(1)}_{k,ref}(n) \equiv 1/(m_\Delta-|k|+1/2) \,\overline{\sum_{ij \in k}} \equiv \bar n^2_{k}$ ($i, j$ are particle indices). The type-2 equivalent based on bin averages is $A^{(2)}_{k,ref}(n) \equiv 1/(m_\Delta - |k|) \sum_{a=1}^{m_\Delta-|k|} \bar n_a \, \bar n_{a+k} \equiv \bar n_k^2$. The type-1 {\em difference} autocorrelation for measure $w$ is defined in terms of sibling (same-event) and mixed pairs as $\Delta A^{(1)}_{k}(w) \equiv 1/(m_\Delta-|k|+1/2) \overline{\sum_{ij \in k} \Delta w_{i} \Delta w_{j}} $.  That form is illustrated in Fig.~\ref{bins} (left panel). We can utilize bin averages $w_{a,ref} \equiv \bar w_a$ to define type-2 difference autocorrelation $\Delta  A^{(2)}_{k}(w) \equiv 1/$\mbox{$(m_\Delta-|k|)$}$ \sum_{a=1}^{m_\Delta - |k|} \overline{\Delta w_{a}\, \Delta w_{a+k}} $. As defined, $\Delta A^{(2)}_{k}(w)$ is a projection by averaging onto its difference variable of binned two-particle {\em covariance} distribution $\overline{\Delta w_a\, \Delta w_b}$ on $(x_1,x_2)$ ($a, b$ are microbin indices). That form is illustrated in Fig.~\ref{bins} (right panel, with $\delta x \rightarrow \Delta x$). The difference autocorrelation, being a projection of $\Delta \rho(x_1,x_2)$ onto its difference variable, is also a covariance distribution (most obvious for the type-2 formulation above), except for the $k=0$ bin which is a variance at the microbin scale. Figs.~\ref{fig3} (right panels) and \ref{fig5} show examples of difference autocorrelations obtained from fluctuation inversion.

\section{The Integral Equation}

We next obtain an exact correspondence between the per-particle variance and a type-2 difference autocorrelation. We use the relation between binning schemes in Fig.~\ref{twobins} to express per-particle variance as the scale integral of a type-1 difference autocorrelation, most efficiently obtained from a direct autocorrelation analysis. We then generalize to 2D momentum subspace $(\eta,\phi)$ to obtain the required integral equation relating fluctuation scale dependence on $(\delta \eta,\delta \phi)$ to joint autocorrelations on axial difference variables $(\eta_\Delta,\phi_\Delta)$.

\subparagraph{Variances and type-2 autocorrelations}

Starting with the definition of total variance we obtain the exact relation of the per-particle variance to a weighted sum of type-2 autocorrelation elements. The total variance as a sum over occupied macrobins is given in Eq.~(\ref{totvar}). Since $\Delta w_s(\delta x) = \sum_{a \in s} \Delta w_a(\epsilon_x)$, the corresponding per-particle variance can be written as a sum over 2D microbins
%\sigma^2_{w/}(m\, \epsilon_x) &=& \frac{1}{\bar n(\delta x)}\, \overline{\sum_{a,b=1}^{m_\delta} \Delta w_a\, \Delta w_b} \\ \nonumber
\bea
\sigma^2_{w/}(\delta x)  &=& \frac{1}{\bar n(\delta x)}\, \sum_{k=1-m_\delta}^{m_\delta-1} \hspace{.05in} \overline{\sum_{a,b=1\cdots m_\delta}^{b=a+k}  \overline{\Delta w_a\, \Delta w_{b}}},
\eea
where the sums over indices $a, b$ include only microbins within macrobins at scale $\delta x$, as in Fig.~\ref{bins} (right panel), and the upper overline indicates an average over all occupied macrobins in the acceptance. Rearranging and noting that $\bar n(\delta x) \equiv m_\delta\, \bar n(\epsilon_x)$ we obtain
\bea \label{type2}
& &\sigma^2_{w/}(\delta x = m_\delta\, \epsilon_x) = \sum_{k=1-m_\delta}^{m_\delta-1} \frac{m_\delta-|k|}{m_\delta}\, \times   \\ \nonumber
&&\hspace{-.1in} \times\,  \overline{\left\{ \frac{\bar n_k(\epsilon_x)}{\bar n(\epsilon_x)}\,\left[ \frac{1}{m_\delta-|k|}\sum_{a,b}^{b=a+k} \frac{\sqrt{{\bar n_a\,\bar n_{b}}}}{\bar n_k(\epsilon_x)}\left( \frac{\overline{\Delta w_a\, \Delta w_{b}}}{\sqrt{{\bar n_a\,\bar n_{b}}}} \right) \right] \right\}}
\eea
where $\bar n_k(\epsilon_x) \equiv 1/ (m_\delta-|k|)\,\overline{\sum_{a,b}^{b=a+k}\sqrt{{\bar n_a\,\bar n_{b}}} }$ is a geometric mean over microbins within macrobins at scale $\delta x$, and the upper overline denotes an average over all macrobins in the acceptance. 

%This relation between per-particle variance and two-particle difference histogram is exact.

The quantities in parentheses are two-particle difference-histogram elements equivalent to density ratio $\epsilon_x \{\Delta \rho(w) /  \sqrt{\rho_{ref}(n)}\}_{ab}$. %defined in the space of Fig.~\ref{bins} (right panel). 
That 2D histogram represent {\em per particle} correlations in measure $w$ on space $(x_1,x_2)$. The quantity in square brackets is a weighted average of those histogram elements over the $k^{th}$ diagonal which {\em estimates} type-2 difference autocorrelation $\Delta \hat A^{(2)}_k(w)/\sqrt{\hat A^{(2)}_{k,ref}(n)}$, the per-particle autocorrelation distribution on $x_\Delta$. 
%In principle, that estimate can depend weakly on macrobin scale through dependence on $m_\delta$ ({\em i.e.,} the fraction of the $k$ diagonal sampled for given scale $\delta x$ -- {\em cf}\, Fig.~\ref{bins}, right panel). 
The quantity in curly brackets estimates autocorrelation elements $\Delta A^{(2)}_k(w)/\sqrt{ A^{(2)}_{k,ref}(n)}$ which would be inferred from inversion of fluctuation scale dependence ({\em cf}\, Eq.~(\ref{xinverse})). If quantity $\bar n_k$ varies significantly (possible on $\eta_\Delta$, depending on collision system, detector acceptance and required precision) the $O(1)$ multiplicity-ratio factor $\bar n_k / \bar n$ may be removed as a systematic correction to obtain the hatted autocorrelation in square brackets. This per-particle relation offers an optimized match between two-particle correlations and fluctuation measures. %In individual applications the per-particle measure format has additional advantages regarding statistical properties and interpretability. 

\subparagraph{Variance differences and type-1 autocorrelations}

We now transition from type-2 to type-1 autocorrelations in order to relate fluctuation scale dependence to the more efficient direct method of obtaining autocorrelations. The quantity in curly brackets in  Eq.~(\ref{type2}) estimates the type-2 difference autocorrelation defined previously as an average over {\em all} microbins on the $k^{th}$ diagonal within the acceptance (Fig.~\ref{bins}, right panel, $\delta x \rightarrow \Delta x$). We rewrite Eq.~(\ref{type2}), replacing the quantity in curly brackets by the difference autocorrelation which it estimates. We then re-express the variance difference in terms of an equivalent expression for the type-1 difference autocorrelation
\bea \label{type1auto}
\sigma^2_{w/}(m_\delta \epsilon_x) &=&    \sum_{k=1-m_\delta}^{m_\delta - 1} \frac{m_\delta-|k|}{m_\delta}  \,\left\{ \frac{\Delta A_k(w) }{ \sqrt{A_{k,ref}(n)}}\right\}^{(2)} \\ \nonumber
%\eea
%\bea
 &=&  2\sum_{k=1}^{m_\delta } \frac{m_\delta-k+1/2}{m_\delta}  \,\left\{ \frac{\Delta A_k(w) }{ \sqrt{A_{k,ref}(n)}}\right\}^{(1)} \\ \nonumber
%\eea
%\bea
&=&  2\sum_{k=1}^{m_\delta} \frac{m_\delta-k+1/2}{m_\delta}\, \times \nonumber  \\ \nonumber
%\eea
%\bea
&& \hspace{-.5in} \hspace{-.1in}\times  \left\{ \frac{\bar n_k(\epsilon_x)}{\bar n(\epsilon_x)}\,\left[ \frac{1}{m_\Delta-k+1/2}\sum_{i,j\in k}^{N}\frac{\overline{\Delta w_i\, \Delta w_{j}}}{\bar n_k(\epsilon_x)} \right] \right\}, \nonumber
\eea
where $\Delta w_i = w_i - \hat w$ for particle $i$, $\hat w$ is the inclusive mean of $w$, $\bar n^2_k(\epsilon_x) \equiv \overline{\sum_{i,j\in k}^{N}} / (m_\Delta-k+1/2)$ for mixed pairs, and the quantity in square bracket is the autocorrelation obtained by direct method (1). Either expression (1) or (2) above can be used to define the integral equation relating fluctuation scale dependence to difference autocorrelations. The two forms correspond to the two binning schemes on the difference variable shown in Fig.~\ref{twobins}.

Inclusive variance $\sigma^2_{\hat w}$ in variance difference $\Delta \sigma^2_{w/}(\delta x)$ corresponds to self pairs in the pair sums of Eq.~(\ref{type1auto}). If those pairs are excluded from two-particle pair distributions and the autocorrelation definitions then the relations above apply to the variance {\em difference}, rather than the scale-dependent variance. For the type-1 autocorrelation we then obtain the integral equation
\bea \label{xinverse}
\Delta \sigma^2_{w/}(m \,\epsilon_x) &=&  2  \sum_{k=1}^{m}  K_{m;k}  \frac{\Delta A_{k}(w;\epsilon_x) }{ \sqrt{A_{k,ref}(n;\epsilon_x)}},
\eea
where $K_{m;k} \equiv (m - k + 1/2)/m$ is an integration kernel representing the macrobin scheme, and we have dropped the subscript $\delta$ from index $m$. Autocorrelation difference density $\Delta \rho$ is related to the autocorrelation difference histogram by $\Delta A_{k}(w;\epsilon_x) \equiv \epsilon^2_x\, \Delta  \rho(w;k\, \epsilon_x)$. Likewise, $A_{k,ref}(n;\epsilon_x) \equiv \epsilon^2_x\, \rho_{ref} (n;k\, \epsilon_x)$. The integral equation is then re-expressed in terms of a density ratio as
\bea
\Delta \sigma^2_{w/}(m\, \epsilon_x) &=&  2  \sum_{k=1}^{m} \epsilon_x  K_{m;k}   \frac{\Delta \rho(w;k\,\epsilon_x) }{ \sqrt{\rho_{ref}(n;k\,\epsilon_x)}}.
\eea
%\frac{\bar n_{k}}{\bar n(\epsilon_x)}

\subparagraph{2D integral equation}

For a 2D scaling analysis on $(\eta,\phi)$ we generalize $\delta x \rightarrow  (\delta \eta,\delta \phi)$ to obtain the per-particle $w$ variance difference (also defining difference factor $\Delta \sigma_{w/}$~\cite{ptprl}) as a 2D discrete integral equation
\bea \label{inverse}
\Delta \sigma^2_{w/}(m \, \epsilon_\eta, n \, \epsilon_\phi) &\equiv& 2 \sigma_{\hat w} \Delta \sigma_{w/}(m \, \epsilon_\eta, n \, \epsilon_\phi) \\ \nonumber
= 4  \sum_{k,l=1}^{m,n} \epsilon_\eta \epsilon_\phi & &\hspace{-.25in} K_{mn;kl}  \,   \frac{\Delta \rho(w;k\,\epsilon_\eta, l\, \epsilon_\phi) }{ \sqrt{\rho_{ref}(n;k\,\epsilon_\eta, l\, \epsilon_\phi)}} ,
\end{eqnarray}
with kernel $K_{mn;kl} \equiv (m - {k + 1/2})/{m} \cdot (n-{l+1/2})/{n}$ representing the 2D macrobin system. This is a Fredholm integral equation which can be inverted to obtain autocorrelation density ratio ${\Delta \rho_{kl}}/{ \sqrt{\rho_{ref,kl}}}$ as a per-particle correlation measure on difference variables $(\eta_\Delta,\phi_\Delta)$ from fluctuation scale dependence of $\Delta \sigma^2_{w/}(\delta \eta,\delta \phi)$.

\section{The Inverse Problem}

We have established the relationship between a variance difference and the integral of an autocorrelation. Given the former as experimental data we wish to infer the latter as conveying the underlying physics in a more interpretable form. Joint (2D) autocorrelation distributions on difference variables $(\eta_\Delta,\phi_\Delta)$ are inferred from fluctuation scale dependence on $(\delta \eta,\delta \phi)$ by inverting those distributions according to Eq.~(\ref{inverse}) (those autocorrelation distributions are by construction symmetric about the origin on $(\eta_\Delta,\phi_\Delta)$). In standard nomenclature the variance difference on the left in Eq.~(\ref{inverse}) is `data' ${\bf D}$ and the autocorrelation density ratio on the right is `image'  ${\bf I}$. Schematically, ${\bf D = {T\, I}} \leftrightarrow {\bf \Delta \sigma^2_{w/}} = {\bf K \, \Delta \rho / \sqrt{\rho}}$.

\subparagraph{Extracting the image} 

Inverting an integral equation from data to image is usually nontrivial and has engendered a large literature~\cite{invtheor}. Integration and inversion are mappings (matrix transformations  ${\bf T}$, ${\bf T}^{-1}$) from one vector space to another. Forward integration from image to data can be represented by ${\bf D = {T\, I} + N}$, (${\bf N}$ represents noise added to the data as part of measurement) with formal inverse $\bf{I' =T}^{-1} \bf {D =  I + T}^{-1} \bf {N } $, where {\bf I} is the true image. The image produced by naive mapping $\bf{I' =T}^{-1} \bf {D}$ is typically dominated by small-wavelength components of image noise $\bf T^{-1}N$, since $\bf T^{-1}$ is a form of differentiation (acting as a `high-pass' filter). A smoothing procedure called {\em regularization} is used to reduce image noise while minimizing distortion of the true image. 

\subparagraph{Regularization}

By {\em smoothing} the inverted image ${\bf I'_\alpha = T}_\alpha^{-1}\, {\bf D}$ with a regularization procedure (smoothing parameter $\alpha$ is defined below) we control the impact on the image of the statistical noise on the data. We treat each bin of the image as a free fitting parameter, but add a smoothing term with Lagrange multiplier  $\alpha$ to the fit $\chi^2$ to obtain $\chi^2_\alpha \equiv {\bf ||D - T\,I'_\alpha||}^2 + \alpha \, {\bf ||L\, I'_\alpha||}^2$. The first term measures the deviation of the integrated  image from the data. The second term, with linear operator {\bf L}, measures the small-wavelength $\lambda$ structure of the image. The optimum choice of $\alpha$ maximally reduces small-$\lambda$ noise on ${\bf I'_\alpha}$ without significantly distorting the signal - the meaningful structure on true image distribution ${\bf I}$. Operator {\bf L} can take various forms~\cite{smooth}. The underlying assumption of a regularization procedure is that the statistical `noise' is sufficiently distinct from the true correlation `signal' on wavelength $\lambda$ that operator {\bf L}, a measure of small-$\lambda$ structure on ${\bf I'_\alpha}$, can distinguish the two. We eliminate most of the small-$\lambda$ noise while retaining as much large-$\lambda$ signal as possible. As $\alpha$ increases, more small-$\lambda$ noise is eliminated. At some point increasing $\alpha$ reduces the signal. By comparing the two elements of $\chi^2_\alpha$ as in Fig.~\ref{fig4} (left panel), the optimum value of $\alpha$ can be determined.

\subparagraph{Statistical and systematic errors}

The autocorrelation (image) error distribution has two components: statistical error which survives smoothing and systematic error due to image distortion by smoothing. The optimum value of $\alpha$ should adequately reduce statistical noise relative to signal while minimally distorting the signal. The residual statistical noise and the signal distortion then depend on the optimum value of $\alpha$. The linear relation between statistical errors on data and image is similar to that for data and image themselves. 
%$\delta^2 {\bf D} = {\bf T}^2\,\delta^2 {\bf  I}$. 
It is straightforward to invert that relation (using the optimized $\alpha$ value) to obtain mean-square errors $\delta^2 {\bf  I}_\alpha$ for the image histogram. Because of the structure of difference autocorrelation density ratio $\Delta \rho / \sqrt{\rho_{ref}}$ those errors tend to be uniform on the difference-variable space, making their characterization simple. 

To estimate smoothing distortions, inferred image ${\bf I}'_\alpha$ can be integrated forward to become a new data distribution, and then inverted again to obtain a twice-smoothed image. That procedure is represented schematically by 1) ${\bf I}'_\alpha = {\bf T}^{-1}_\alpha {\bf D}$ (smoothed image from data inversion), 2) ${\bf D}_\alpha = {\bf T\,I}'_\alpha$ (forward-integrated smoothed image), 3) ${\bf I}'_{\alpha\, \alpha} = {\bf T}^{-1}_\alpha {\bf D}_\alpha$ (twice-smoothed image). By comparing image distributions ${\bf I}'_{\alpha}$ and ${\bf I}'_{\alpha\,\alpha}$, assuming that the smoothing process is linear, one can estimate the smoothing distortions, usually largest at points of maximum gradient on the image. The smoothing error can  then be displayed as a histogram or summarized as a fractional error of peaked structure on the image distribution.

\section{Monte Carlo Examples}

We apply the inversion method to Monte Carlo data for three reasons: 1) these Monte Carlos produce event ensembles with fairly simple correlation content which are amenable to the type of fluctuation/correlation analysis we wish to apply to nuclear collision data, 2) inversion of Hijing heavy ion fluctuations illustrates how the inversion process leads to simple and direct interpretation of fluctuations in terms of physical mechanisms and 3) comparison of the Pythia fluctuation inversion with a direct computation of Pythia correlations (easily done for p-p collisions) confirms that the inversion process precisely reveals the underlying autocorrelation with negligible distortion. Those basic demonstration goals could be accomplished as well with a simpler physics Monte Carlo, or with a Monte Carlo having no physical correspondence to data ({\em e.g.,} the gaussian on a flat background). The point of this exercise is to make controlled comparisons between inversion and direct pair-counting analysis to establish the precision of the fluctuation inversion, and to perform a controlled inversion study of a Monte Carlo with known physics content for calibration, before applying the method to real data.

\subparagraph{Hijing Au-Au 200 GeV $p_t$ correlations}

Event generator Hijing-1.37~\cite{hijmc} was used to produce a total of 1M minimum-bias Monte Carlo collision events of two types: 1) {\em quench-off}\, Hijing -- jet production enabled but no jet quenching  and 2) {\em default} or {\em quench-on} Hijing -- jet production and jet quenching both enabled. Each event type was then split into six centrality classes consisting of roughly equal fractions of total cross section. A more complete analysis of Hijing fluctuation data and inversion to $p_t$ autocorrelations is described in~\cite{hijsca}. For the fluctuation analysis charged particles with pseudorapidity $|\eta| < $ 1, transverse momentum $p_t \in [0.15,2]$ GeV/c and full azimuth were accepted. Fluctuation scale dependence was then determined using the variance difference defined in Eq.~(\ref{EqA11}), with $w \rightarrow (p_t - n\, \hat p_t)$ and $w/ \rightarrow p_t:n$. 

%%%%%%%%%%%%%%%%%%%%%%%%%%%%%%%
\begin{figure}[h]
\begin{tabular}{cc}
\begin{minipage}{.47\linewidth}
\includegraphics[keepaspectratio,width=1.65in]{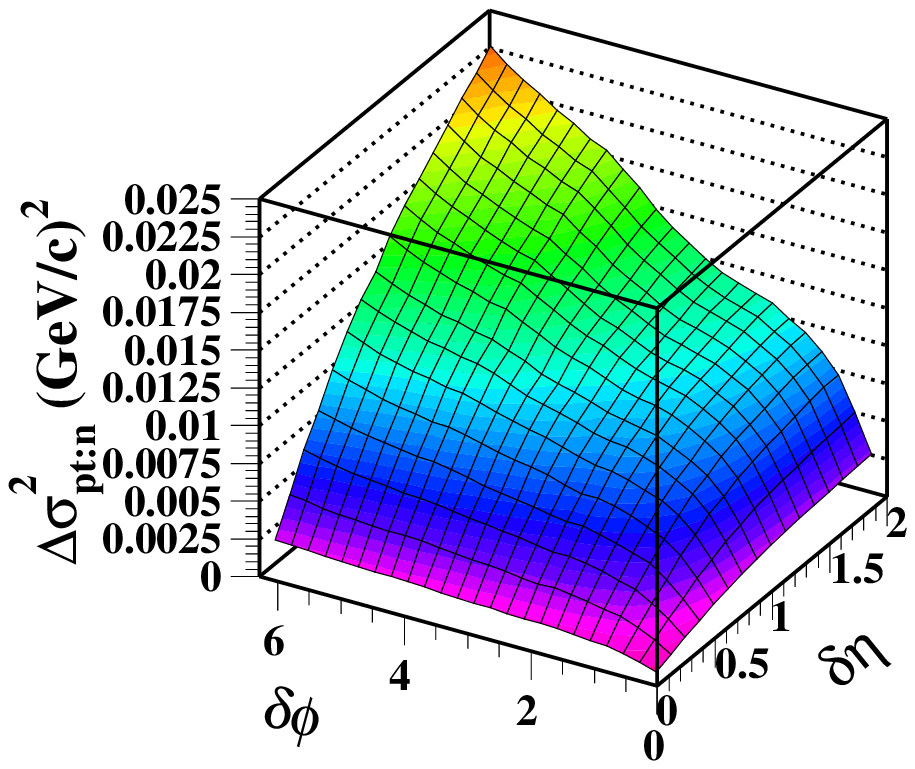}
\end{minipage} &
\begin{minipage}{.47\linewidth}
\includegraphics[keepaspectratio,width=1.65in]{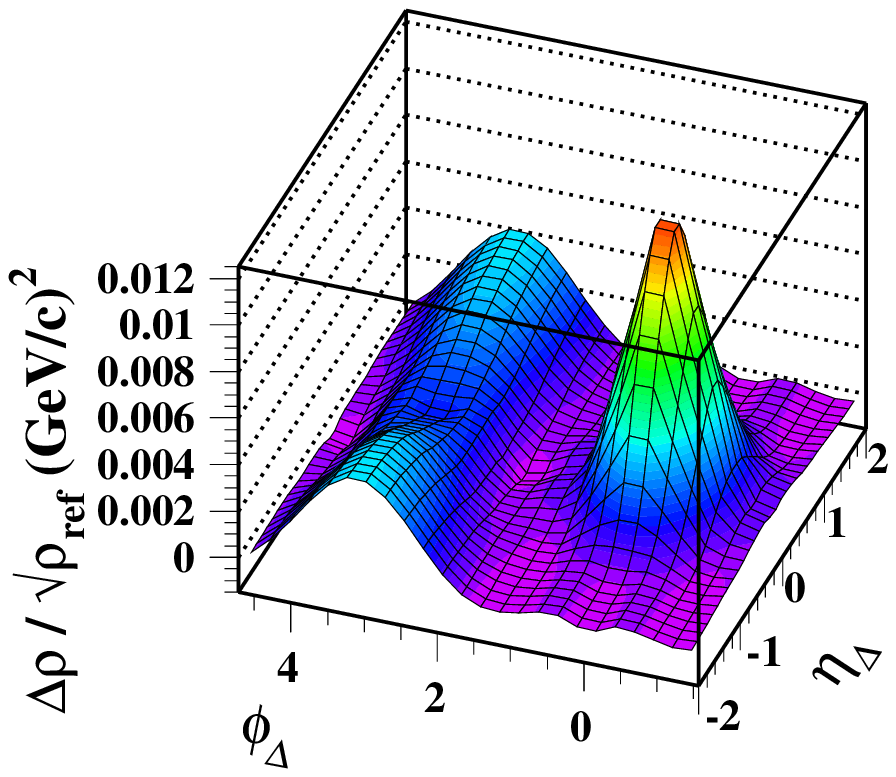}
\end{minipage}\\
\begin{minipage}{.47\linewidth}
\includegraphics[keepaspectratio,width=1.65in]{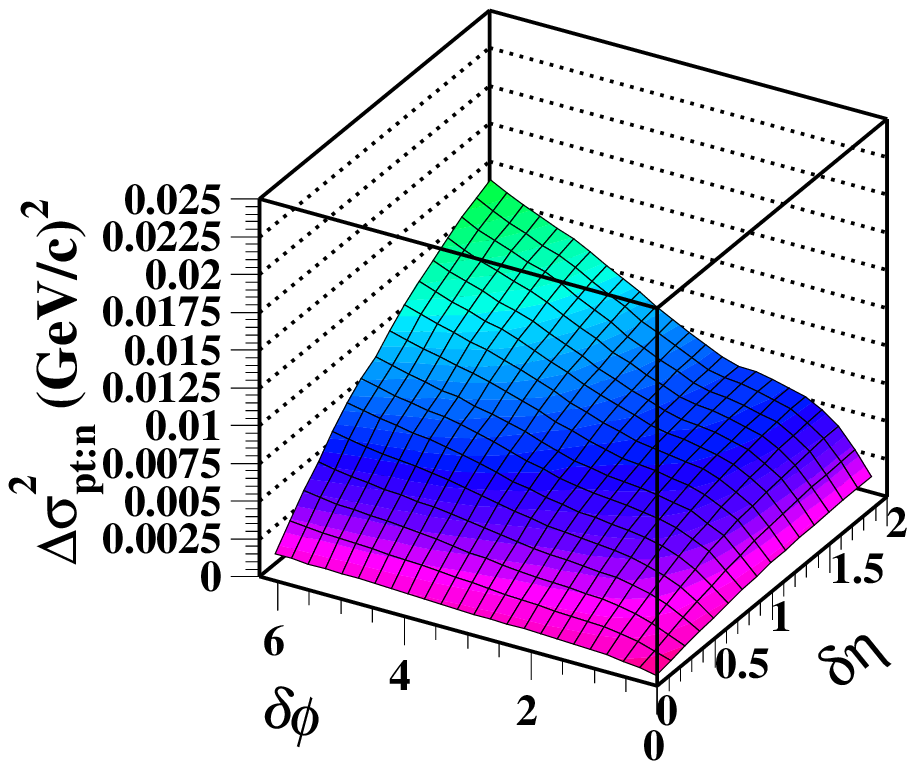}
\end{minipage} &
\begin{minipage}{.47\linewidth}
\includegraphics[keepaspectratio,width=1.65in]{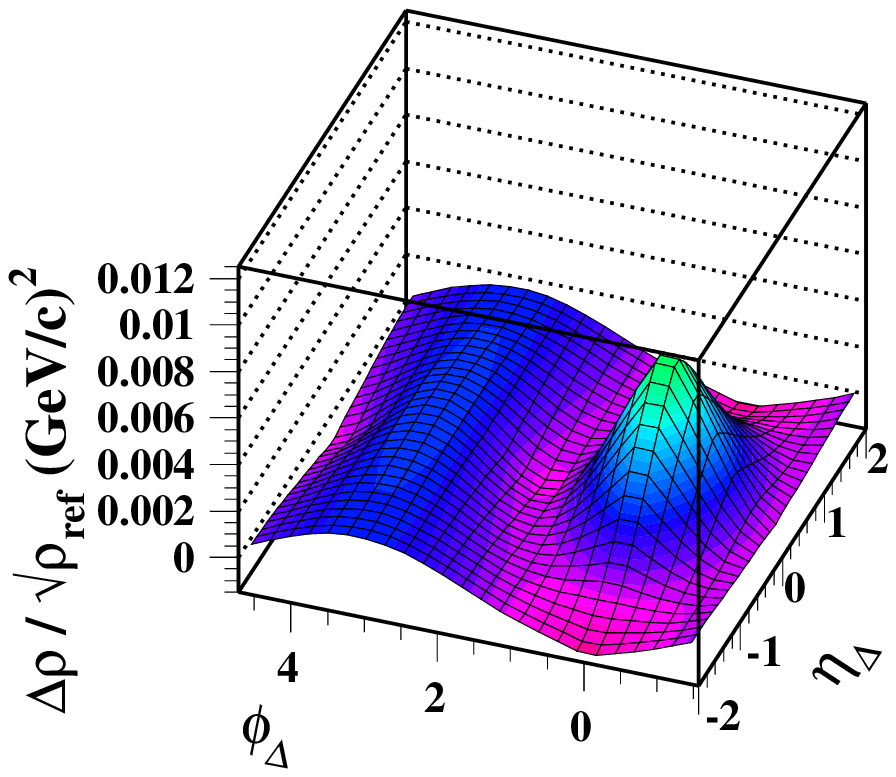}
\end{minipage}\\
\end{tabular}
\caption{Left panels: $\Delta \sigma^2_{p_{t}:n}$ (GeV/c)$^2$ distributions on binning scales $(\delta \eta,\delta \phi$) for central 0-15\% collisions and two Hijing states: quench off (top panel) and quench on (bottom panel). Right panels: Corresponding autocorrelations on difference variables ($\eta_{\Delta},\phi_{\Delta}$). 
\label{fig3}}
\end{figure}
%%%%%%%%%%%%%%%%%%%%%%

Fig.~\ref{fig3} shows fluctuation scale dependence and corresponding autocorrelations for 0-15\% central collisions and two Hijing configurations: quench-off (top panels) and quench-on (bottom panels). In the left panels are variance differences $\Delta \sigma^2_{p_t:n}(\delta \eta,\delta \phi)$ [units (GeV/c)$^2$], which typically increase monotonically with pseudorapidity scale $(\delta\eta)$ but are more complex on azimuth scale $(\delta\phi$). 
%While fluctuation measurements can certainly be compared to theoretical predictions, underlying physical mechanisms are often more evident in the corresponding autocorrelation distributions. 
%Whereas the physical interpretation of such fluctuation distributions is typically ambiguous, interpretation of the corresponding autocorrelations is often straightforward and gives access to critical parameters of physical phenomena, in this case the peak amplitude and widths. 
Autocorrelation distributions $\Delta \rho / \sqrt{\rho_{ref}}$ [also units (GeV/c)$^2$] on $(\eta_\Delta,\phi_\Delta)$ shown in Fig.~\ref{fig3} (right panels) contain two features: a same-side component ($|\phi_\Delta| < \pi/2$) and an away-side component  ($\pi/2 < |\phi_\Delta| < \pi$). The dominant same-side peak can be identified with {\em minimum-bias} parton fragments (minijets): no high-$p_t$ trigger condition is imposed in the analysis~\cite{hijsca}. 

%%%%%%%%%%%%%%%%%%%%%%%%%%%%%%%
\begin{figure}[h]
\begin{tabular}{cc}
\begin{minipage}{.47\linewidth}
\includegraphics[keepaspectratio,width=1.65in]{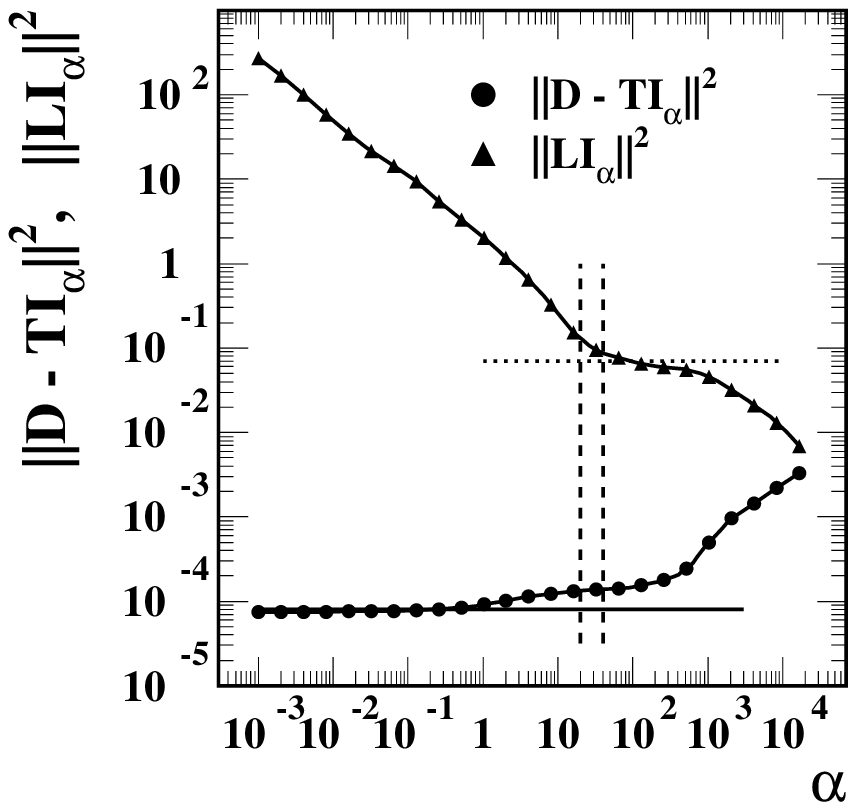}
\end{minipage} &
\begin{minipage}{.47\linewidth}
\includegraphics[keepaspectratio,width=1.65in]{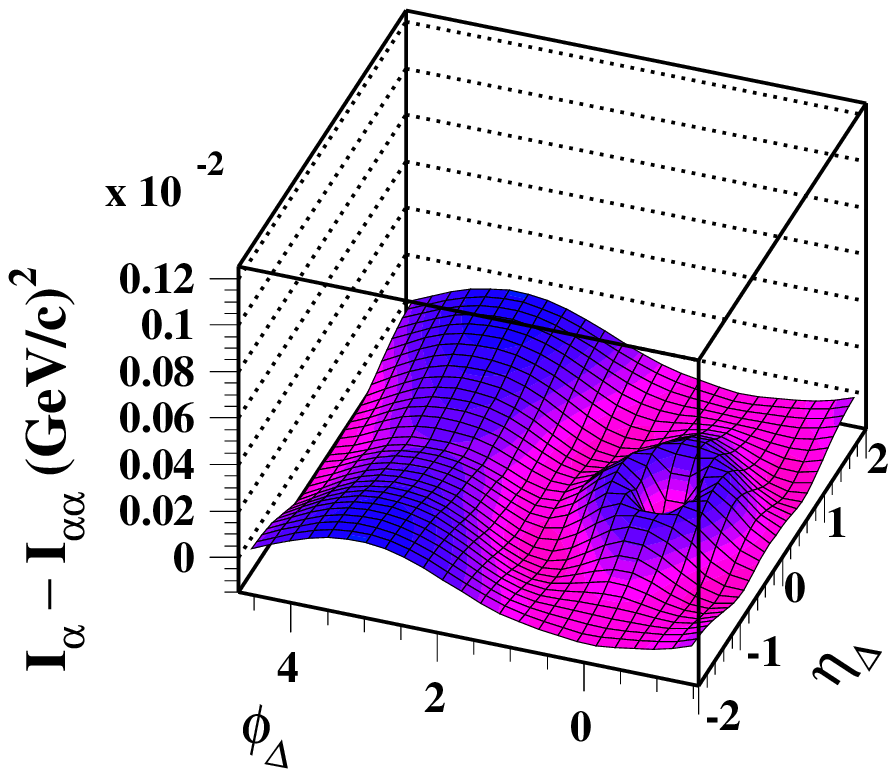}
\end{minipage}\\
\end{tabular}
\caption{Left panel: Distribution of $||{\bf D - T I}_\alpha||^2$ (dots) and $||{\bf L I}_\alpha||^2$ (triangles) $vs$ $\alpha$ for the 0-15\% central quench-off data in Fig.~\ref{fig3} (upper-right panel). Deviation of $||{\bf D - T I}_\alpha||^2$ from the extrapolation (horizontal solid line) indicates systematic error due to image smoothing. The vertical dashed lines bracket the optimal choice of $\alpha$. Right panel: Distribution of estimated smoothing error ${\bf I}_{\alpha} - {\bf I}_{\alpha\alpha}$ on $(\eta_\Delta,\phi_\Delta)$, also for the 0-15\% central quench-off autocorrelation in Fig.~\ref{fig3} (upper-right panel). Note the 10$\times $ scale reduction compared to Fig.~\ref{fig3}. \label{fig4}}
\end{figure}
%%%%%%%%%%%%%%%%%%%%%%

An illustration of $\alpha$ optimization is given in Fig.~\ref{fig4}. Distributions $||{\bf D - T I}_\alpha||^2$ and $||{\bf L I}_\alpha||^2$  on $\alpha$ shown in Fig.~\ref{fig4} (left panel) can be used to determine the optimum $\alpha$ value. The $||{\bf D - T I}_\alpha||^2$ distribution is nearly constant at smaller $\alpha$, indicating no loss of valid structure by smoothing. The  $||{\bf L I}_\alpha||^2$ distribution falls rapidly, indicating substantial noise reduction. At some value of $\alpha$\, $||{\bf D - T I}_\alpha||^2$ begins to rise from the constant (extrapolated by the solid line), the deviation representing information loss from the image. Beyond that point there is a tradeoff between smoothing distortion and noise reduction. In the example of Fig.~\ref{fig4} the noise exhibits a stationary point at the same position (upper dotted line). The optimum choice of $\alpha$ is in the region bracketed by the vertical dashed lines.

Autocorrelation errors have two components: statistical errors derived from corresponding errors in fluctuation data, and distortion errors due to smoothing. Statistical errors on the image are found by inverting statistical errors on fluctuation data. The structure of this error distribution is usually simple, and can be characterized by a single number for each autocorrelation. Smoothing errors are estimated by passing inferred image ${\bf I}_\alpha$ a second time through the integration/inversion process to obtain ${\bf I}_{\alpha\alpha}$. Difference ${\bf I}_\alpha - {\bf I}_{\alpha\alpha}$ provides an upper limit on the smoothing distortion (some of the change includes further reduction of statistical error). A typical image-smoothing error distribution is shown in Fig.~\ref{fig4} (right panel). The per-bin systematic error on autocorrelation distributions is typically a few percent of maximum autocorrelation values, somewhat larger peak errors lying within small regions corresponding to large image gradients.

%Summary of inversion process and error estimation for XXX million Hijing quench-off, 0-5\% central collisions. In four upper panels starting from top left: $\Delta \sigma_{p_{t}:n}$ data distribution on scale $(\delta \eta,\delta \phi$), corresponding image autocorrelation on difference variables ($\eta_{\Delta},\phi_{\Delta}$) for selected smoothing paramter $\alpha$, forward integration of that image back to the data space, and second inversion of the integrated image with the same smoothing parameter. At bottom left and right are the corresponding difference histograms from the upper two panels revealing distributions of estimated systematic and residual statistical error on data and image distributions respectively.

These Hijing results illustrate the process of reducing event-wise fluctuations to autocorrelations and their physical interpretation. The autocorrelations revealed in this study and in~\cite{hijsca} have features consistent with the physical model of jet production and quenching in that Monte Carlo. They represents the first determination of {\em transverse momentum} correlations from $\langle p_t \rangle$ fluctuations for any nuclear collision data or model. Transverse momentum autocorrelations provide a much more critical and differential test of the Hijing model than was previously conducted (the model fails those comparisons with data). The inversion process clearly reveals what mechanism in Hijing generates $\langle p_t \rangle$ fluctuations, which was previously a matter of speculation. These jet correlations were not obtained with a jet trigger {\em per se} (leading-particle or jet-cone analysis), but instead result from an {\em unbiased}, symmetric two-particle correlation analysis.

\subparagraph{Pythia p-p 200 GeV $n$ correlations}

\begin{figure}[h]
\includegraphics[keepaspectratio,width=1.65in]{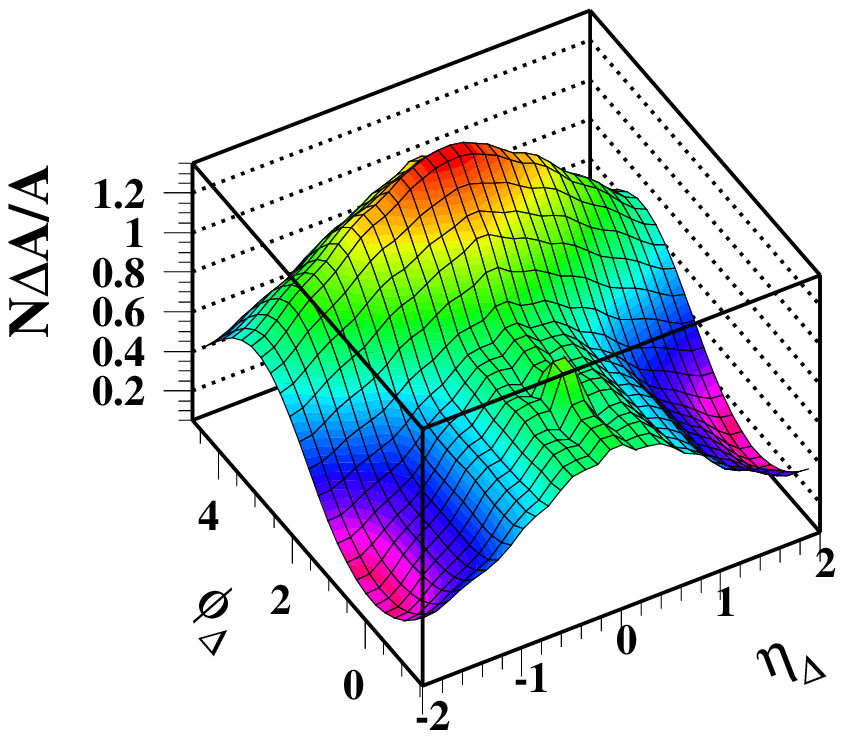}
\includegraphics[keepaspectratio,width=1.65in]{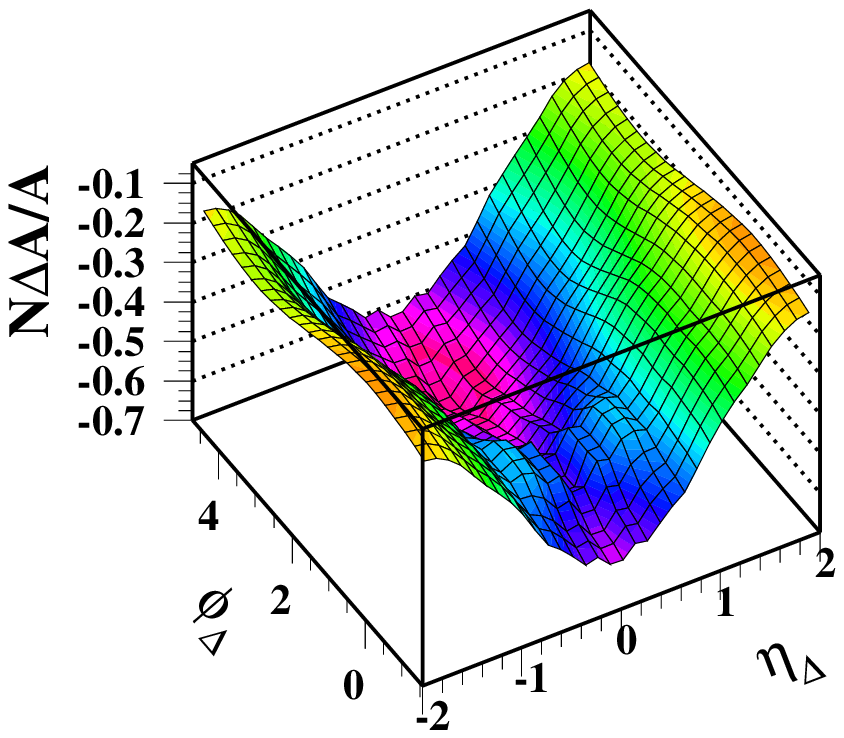}
\includegraphics[keepaspectratio,width=1.65in]{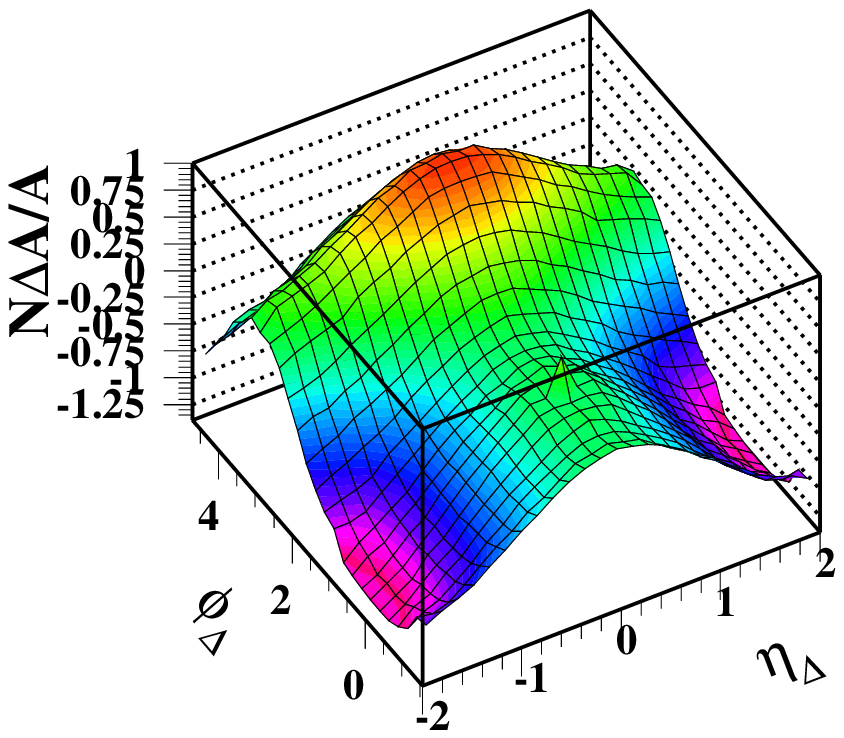}
\includegraphics[keepaspectratio,width=1.65in]{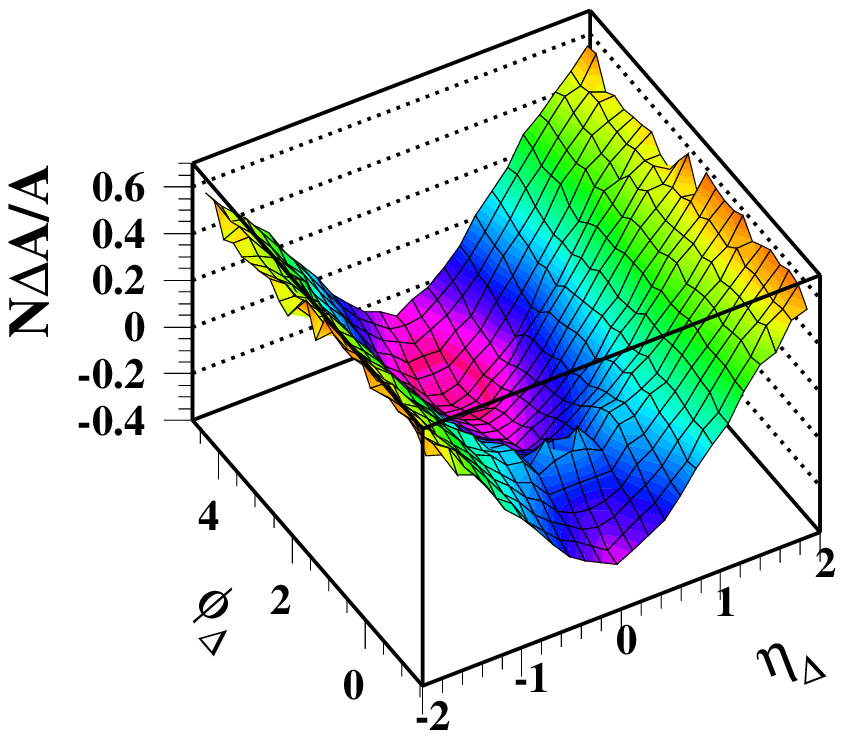}
\caption{Pair-number autocorrelations on difference variables $(\eta_\Delta,\phi_\Delta)$ for p-p collisions at 200 GeV from Pythia-6.131 obtained from fluctuation inversion (upper panels) and from direct pair counting (lower panels) for isoscalar (left panels) and isovector (right panels) autocorrelations. \label{fig5}}
\end{figure}

The Pythia-6.131 Monte Carlo~\cite{pythia} was used to provide a high-statistics sample of 200 GeV p-p collision events. Analysis of those data provides a demonstration of the precision possible with fluctuation inversion and a check on the algebraic system of this paper. We make a comparison between autocorrelations determined directly according to Eq.~(\ref{auto1}) and indirectly by fluctuation inversion according to Eq.~(\ref{inverse}). In Fig.~\ref{fig5} are charge-independent (left panel, like-sign $+$ unlike-sign pair combinations) and charge-dependent (right panel, like-sign $-$ unlike-sign pair combinations) pair-number autocorrelations on difference variables $(\eta_\Delta,\phi_\Delta)$~\cite{axialcd}. The autocorrelations were obtained by fluctuation inversion (upper panels) and directly from binned pair numbers (lower panels). Detailed comparison of the shapes of each pair of distributions indicates that fluctuation inversion is a very precise method of determining autocorrelations.

%%%%%%%%%%%%%%%%%%%%%%%%%%%%%%%
%%%%%%%%%%%%%%%%%%%%%%

The shift of vertical scale for upper and lower panels in Fig.~\ref{fig5} is due to the choice of sibling {\em vs} mixed pair normalization used in the direct determination of pair-number autocorrelations. Pair-number autocorrelations are constructed from pair sums $\hat n_{ij} \equiv n_{ij} / \sum_{ij} n_{ij}$ {\em normalized to unit integral} for sibling and mixed pairs~\cite{axialcd}. Normalized pair-ratio histograms $\hat r_{ij} \equiv \hat n_{ij,obj} / \hat n_{ij,ref}$ are first formed on microbinned spaces $(\eta_1,\eta_2)$ and $(\phi_1,\phi_2)$. Those distributions are then projected to form type-2 autocorrelation ratios $\Delta A^{(2)}_k / A_{k,ref} \equiv \hat r_k - 1$ on $(\eta_\Delta,\phi_\Delta)$. The pair-number difference autocorrelation $\Delta A^{(2)}_k$ so constructed has zero integral, and therefore always varies about zero as in the lower panels of Fig.~\ref{fig5}. Autocorrelations from fluctuation inversion retain their absolute vertical position: the offset missing from the zero-integral type-2 autocorrelation is equivalent to a fluctuation measurement at the acceptance scale. The factor $\bar N$ in Fig.~\ref{fig5} defines a transitional form: $\bar N \Delta A^{(2)}_k / A_{k,ref} \simeq \Delta \eta \Delta \phi \cdot \Delta \rho / \sqrt{\rho_{ref}}$

The fluctuation inversion procedure is a general analysis technique. It is not unique to these Monte Carlos or to nuclear physics. It can be applied to any problem in which correlated distributions are produced as an event sequence and there is persistent event-wise structure over an event ensemble. 

%d^2\bar N /d\eta/d\phi \, \Delta \rho^{(2)}_k / \rho_{k,ref} \rightarrow m_\Delta n_\Delta \,\epsilon_\eta \epsilon_\phi\, \sqrt{\rho_{k,ref}}\,\Delta \rho^{(2)}_k / \rho_{k,ref}$. 
%Thus, quantity $\bar N \Delta A / A_{ref}$ plotted in Fig.~\ref{fig5} is approximately related to quantity $\Delta \rho / \sqrt{\rho_{ref}}$ by factor $\Delta \eta \Delta \phi = 2 \times 2\pi$.

\section{Summary}

Event-wise fluctuations in the hadronic final state of in nuclear collisions, a manifestation of multiparticle correlations produced by collision dynamics, are relatively easy to calculate but often difficult to interpret. Autocorrelations provide a powerful method for direct study of multiparticle correlations but are computationally expensive when constructed directly from pair ratios. For relativistic collisions, autocorrelation projections to subspaces of reduced dimensionality retain most of the correlation information of the full two-particle momentum space in a more compact form. 

In this paper we have derived the exact relation between autocorrelations and fluctuation scale dependence in the form of an integral equation. We have introduced the method required to invert that equation, and the procedure used to solve the technical problem of regularizing statistical noise. Using Monte Carlo simulations we have made detailed comparisons of autocorrelations determined directly from pair ratios and from fluctuation inversion with regularization from  the same particle data to illustrate the precision of the method.

The analysis of the Hijing Monte Carlo illustrates how inversion of fluctuations to autocorrelations permits direct interpretation in terms of physical mechanisms, in this case angular correlations of transverse momentum from jet fragmentation. The inversion method applied to transverse momentum fluctuations provides the first access to velocity correlations in nuclear collisions. This procedure is not unique to nuclear physics data, but can be applied to any point sets arranged in event ensembles. 

%%%%%%%%%%%%%%%%

%We have also described a procedure to determine statistical and systematic errors of the resulting autocorrelations, in particular distortions due to regularization. 

%-------------------------------------------------------------------

\end{document}